# Towards the theory of hardness of materials.

## Artem R. Oganov[1,2], Andriy O. Lyakhov[1]

[1] *Department of Geosciences, Department of Physics and Astronomy, and New York Center for Computational Sciences, Stony Brook University, Stony Brook, New York 11794-2100, USA*

[2] *Geology Department, Moscow State University, 119992 Moscow, Russia*

**Recent studies showed that hardness, a complex property, can be calculated using very simple approaches or even analytical formulae. These form the basis for evaluating controversial experimental results (as we illustrate for TiO$_2$-cotunnite) and enable a systematic search for novel hard materials, for instance, using global optimization algorithms (as we show on the example of SiO$_2$ polymorphs).**

We all know that diamond is very hard, while graphite and talc are soft. Hardness, as a property of materials, determines many of their technological applications, but remains a poorly understood property. The aim of this Special Issue is to review some of the most important recent developments in the understanding of hardness.

Mohs's relative scale of hardness appeared in the XIX century and is still widely used by mineralogists; in this scale talc has hardness 1, and diamond has hardness 10. There are several absolute definitions of hardness – the most popular ones being the Knoop and Vickers tests of hardness, which involve indentation (rather than scratching, as in Mohs's hardness). The absolute hardness is measured in GPa, the same units as pressure or elastic moduli (bulk modulus, shear modulus). This hints that hardness may be correlated with the elastic properties – indeed, there is such a correlation, especially with the shear modulus. However, hardness is obviously a much more complex property than elasticity, as it involves also plastic deformation and brittle failure. For these reasons, a complete picture of hardness cannot be given only by the ideal crystal structure and its properties, but must include also defects (in particular, dislocations) and grain size. The latter is related to hardness through a particularly important phenomenon, known as the Hall-Petch effect – hardness increases as the particle size decreases (in reality, there is a maximum in the nanometer range). Thus, it is possible to significantly boost a material's hardness by creating nanoparticle aggregates and nanocomposites: while the hardness of diamond single crystals varies between 60-120 GPa depending on the direction [1], nanodiamond turns out to be much harder, with the isotropic hardness of up to 120-140 GPa [2]. Cubic BN has Vickers hardness of 40-60 GPa in bulk crystals, but its nanocomposites are almost as hard as diamond, with the Vickers hardness of 85 GPa [3]. Hardness of nanomaterials is discussed in reviews of Tse [4] and Gao [5] in this Issue.

With the detailed understanding of hardness still elusive, can one invent a practical recipe for predicting the hardness of a material on the basis of its crystal structure? This would mean ignoring dislocations and grain boundaries, which is fundamentally incorrect – but a number of practical recipes, invented recently, turn out to give reasonable results, certain predictive power and great fundamental value. These include the ideal strength

(which often attains values in surprisingly good agreement with experimental hardnesses – see [4]) and analytical models [6-10], which hold a potential of revolutionizing the field of superhard materials. These are focus of this Issue. Differing in mathematical and also somewhat in physical details, analytical models have much in common; hardness is high when:

- (i) the average bond strength is high
- (ii) the number of bonds per unit volume is high
- (iii) the average number of valence electrons per atom is high
- (iv) bonds are strongly directional (i.e. have a large covalent component) – ionicity and metallicity decrease hardness

The requirement of high bond strength indicates that compounds light elements, which form extremely strong and short bonds, are particularly promising; some transition metals (e.g., W, Ta, Mo, Re) can also form very strong (although not quite as directional) bonds and have a high number of valence electrons and their compounds should also be carefully examined.

Diamond, a dense phase with strong and fully covalent bonds, satisfies all of the conditions (i)-(iv). Cubic BN, with partially ionic bonds, has a somewhat lower hardness. Graphite, though containing stronger bonds than in diamond, has a much lower number of atoms and bonds per unit volume, and must therefore be softer[1]. Cold compression of graphite [11] resulted in a peculiar superhard phase, the structure of which has been understood only recently [12] and has a much greater density and lower anisotropy than graphite.

The requirement of high bond density means that often superhard materials will have to be synthesized at high pressure – this is the case of diamond [13,14], cubic BN [15], cubic $BC_2N$ [16] and BCN [17], boron-enriched diamond with approximate composition $BC_5$ [18], and novel partially ionic phase of elemental boron, $\gamma$-$B_{28}$ [19,20]. All the listed materials can be decompressed to ambient conditions as metastable phases, but this is hardly a limitation to their performance. Much more critical is the fact that to be practically useful, the material should be synthesizable at pressures not higher than ~10 GPa, because at higher pressures synthesis can be dome only in tiny volumes (except in shock-wave synthesis, which may be a viable route for useful materials at ultrahigh pressures). High-pressure studies of materials are often tricky, and the field of high-pressure research is full of both exciting discoveries and misdiscoveries. For instance, it has been claimed [21] that $TiO_2$-cotunnite, quenched from high pressure, is the hardest known oxide with the Vickers hardness of 38 GPa. While it is hard to experimentally appraise such results obtained on tiny samples, theoretical models can help to distinguish facts from artifacts: within any of the models presented in this Issue, the hardness of $TiO_2$-cotunnite varies in the range 7-20 GPa, our preferred and perhaps the most reliable result (based on extended model [8]) being 15.9 GPa, i.e. this material is certainly about as soft as common quartz (whose Vickers hardness is 12 GPa) and softer than common corundum, $Al_2O_3$ (21 GPa), or stishovite, $SiO_2$ (33 GPa) [22], or $B_6O$ (45 GPa) [23]. Such relatively low hardness of $TiO_2$-cotunnite is only natural, given its very large bond ionicity and high coordination number of Ti (ninefold), i.e. relatively weak and non-directional Ti-O bonds. Thus, ultrahard $TiO_2$-cotunnite is a clear artifact.

---

[1] For materials like graphite, it is essential to take anisotropy into account; the hardness models presented in this Special Issue are isotropic (i.e. give a single overall value of hardness for the material), though an anisotropic extension has just been developed [10].

Thanks to such models of hardness it is now possible to systematically search for superhard materials. We performed a search for the hardest structure of $SiO_2$ at atmospheric pressure, by combining the evolutionary global optimization algorithm USPEX [24-27] with an extended version of the hardness model [8]. To enable very fast exploratory calculations, we relaxed all trial structures using a simple interatomic potential based on the model of Sanders et al. [28] using the GULP code [29]. Hardness was evaluated on fully relaxed structures. Fig. 1 shows how the computed hardness evolved during the simulation (shown here for the system with 24 atoms in the unit cell, though we also explored other system sizes), and it is clear how harder and harder structures are found as the run progresses. The four hardest structures found in all runs (Fig. 2) turned out to be (i-ii) two well-known phases stishovite (rutile-type structure) and seifertite ($\alpha$-$PbO_2$-type structure), (iii) a 3x3 kinked-chain structure, intermediate between stishovite and seifertite, and (iv) a cuprite-type phase (cuprite-type $SiO_2$ is hitherto unknown, but cuprite-type ice X is the densest known phase of ice [32] – and ice phases have strong structural similarities with tetrahedral silica polymorphs). We re-relaxed these structures with more accurate DFT calculations using the generalized gradient approximation [33] and the VASP code [34], and the recomputed hardnesses were 28.9 GPa for stishovite (close to the experimental hardness of 33 GPa [22]), 29.6 GPa for seifertite, 29.3 GPa for the 3x3-structure, and 29.5 GPa for $SiO_2$-cuprite. All these theoretical hardnesses are underestimates, due to the known modest overestimation of bond lengths in the functional [33] and are extremely close to each other. However, it is likely that seifertite, being the densest phase of silica known at 1 atm, is also the hardest one. Seifertite has a stability field on the phase diagram at high pressures (see [35] for discussion) and is quenchable to ambient conditions.

This simple test shows that systematic prediction and design of new ultrahard and superhard materials is now possible, and a central role in this new direction of research is played by the simple – yet powerful – models of hardness, which are the main topic of this Special Issue. Reviews by Gao and Gao [5], Tse [4], and Mukhanov et al. [36] discuss various theoretical models of hardness, while Li et al. [37] focus on search for novel superhard crystal structures. Reviews of Mukhanov et al. [36] and Shirai [38] discuss in details a particularly important and interesting case of boron-rich solids, for which traditional models of hardness have difficulties due to the essential role of multicenter bonding (see also [39] for an exciting discussion). While many further developments are still needed in the theory of hardness, the current state of this field, reviewed in this Special Issue, shows its great utility and promise.

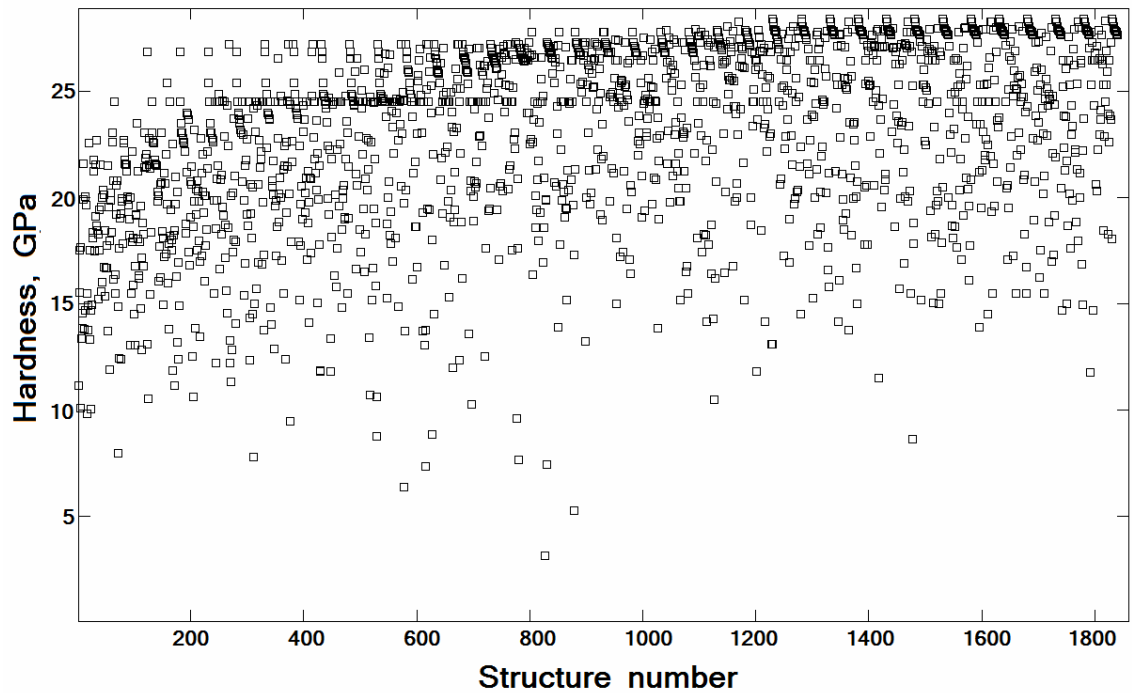

**Fig. 1.** Evolution of the theoretical hardness in an evolutionary global optimization run for $SiO_2$ with 24 atoms in the unit cell.

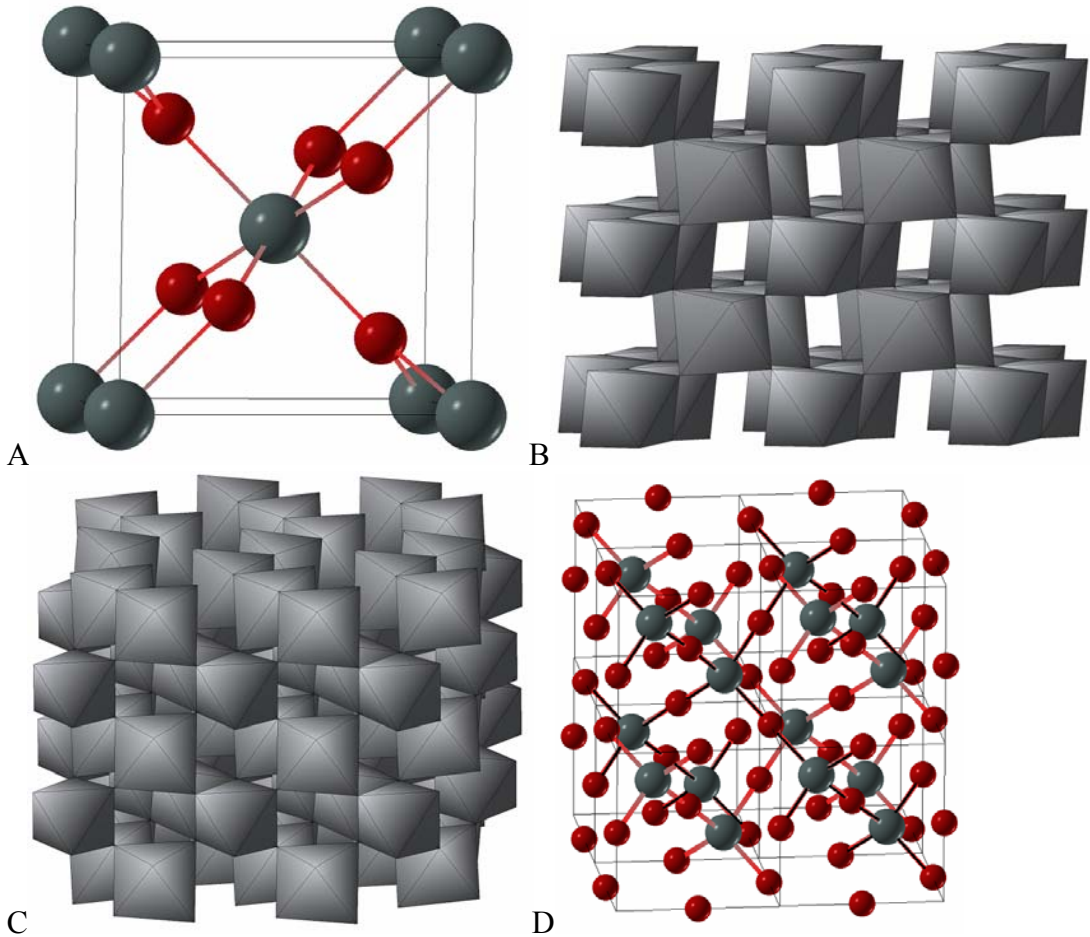

**Fig. 2.** Structures of stishovite (a), seifertite (b), 3x3 phase (c), and cuprite-type (d) modifications of SiO$_2$. Close-packed structures related to (a-c) were discussed in [30]. The cubic cuprite structure (space group *Pn3m*) has lattice parameter a=3.80 Å and the following atomic positions: Si (0.75,0.75,0.75) and O (0,0.5,0). The cuprite structure has two interpenetrating cristobalite-type (or diamond-like) structures (it is a "3D-catenane", as P.M. Zorkii christened it by analogy with interlocked catenane molecules) – not surprisingly, this structure is almost twice (1.88 times) as dense as high cristobalite (it is also 1.45 time denser than quartz, and 1.14 times less dense than stishovite and ). Hypothetical cuprite-type BeF$_2$*SiO$_2$ was already suggested [31] to be a very hard material. Our computed bulk modulus of SiO$_2$-cuprite is 276 GPa, its pressure derivative *Ko*'=6.4. This phase is 0.38 eV/atom less stable than quartz, and 0.2 eV/atom less stable than stishovite.

**Acknowledgments.** This research is funded by the Research Foundation of Stony Brook University, Intel Corporation, and Rosnauka (Russia, contract 02.740.11.5102). Calculations were performed on the NYBlue supercomputer (NYCCS), on the Skif MSU supercomputer (Moscow State University, Russia), and at the Joint Supercomputer Center (Russian Academy of Sciences). A.R.O. would like to thank H. Wang, and V.L. Solozhenko for many scientific discussions, and X.-F. Zhou for discussions of the ideal strength and its relationship with hardness. As the Editor of this Special Issue, A.R.O. would like to thank all authors who contributed to it.